\documentclass[twocolumn]{revtex4-1}

\pdfoutput=1 
\usepackage{graphics} 
\usepackage{graphicx}

\begin{document}

\title{Direct calculation of the solid-liquid Gibbs free energy difference\\in a single equilibrium simulation}


\author{Ulf R. Pedersen} 
\email{ulf.pedersen@tuwien.ac.at} 
\affiliation{Institute of Theoretical Physics, Vienna University of Technology, Wiedner Hauptstrasse 8-10, A-1040 Vienna, Austria} 
\affiliation{Faculty of Physics, University of Vienna, Boltzmanngasse 5, A-1090 Vienna, Austria} 

\date{\today}

\begin{abstract}
Computing phase diagrams of model systems is an essential part of computational condensed matter physics.
In this paper we discuss in detail the interface pinning (IP) method for calculation of the Gibbs free energy difference between a solid and a liquid. This is done in a single equilibrium simulation by applying a harmonic field that biases the system towards two-phase configurations. The Gibbs free energy difference between the phases is determined from the average force that the applied field exerts on the system. As a test system we study the Lennard-Jones model. It is shown that the coexistence line can be computed efficiently to a high precision when the IP method is combined with the Newton-Raphson method for finding roots. Statistical and systematic errors are investigated. Advantages and drawbacks of the IP method are discussed. The high pressure part of the temperature-density coexistence region is outlined by isomorphs.
\end{abstract}

\maketitle

\section{Introduction} 
An important aspect of computational condensed matter physics is to compute phase diagrams of model systems. The naive approach is to preform a long-time simulation at a selected state point and hope that the system by itself finds its preferred phase, i.e. the phase with the lowest Gibbs free energy. In most cases this strategy is not viable since first-order transitions are associated with hysteresis. Thus the system is likely to be stuck in a metastable phase. The origin of this hysteresis effect is the formation of an interface between two phases. The surface tension of the interface gives rise to a free energy barrier that the system has to overcome to transform from one phase to the other \cite{becker1935}. A conceptual appealing and widely used strategy to overcome the hysteresis problem is to preform simulations starting from an initial configuration with two phases in a periodic box \cite{ladd1977,landman1986,mori1995,kyrlidis1995,agrawal2003,morris2002,hoyt2002,sibugaga2002,fernandez2006,vega2008,
weingarten2009,timan2010,pedersen2011_lwotp}. When a steady state situation is 
reached the stable phase will grow at the expense of the other phase. The disadvantages of this approach 
are that: i) it relates to a non-equilibrium 
computation that cannot be done {\it ad infinitum}; ii) a sufficiently large system is needed to reach the steady state growth; iii) thermal fluctuation will result in some probability that the system will move towards the disfavored phase. In a recent paper \cite{pedersen_short} these problems were resolved by introducing the so-called ``interface pinning'' (IP) method. In short, the idea of this method is to compute the average force needed to keep the system in the two-phases state. This is done by connecting the system to a harmonic field which couples to an order-parameter that distinguished between the two phases of interest. The Gibbs free energy difference between the phases is determined by the average force that the applied field exerts on the system. Thus, a standard {\it ad infinitum} equilibrium simulation gives the information needed to computed the Gibbs free energy difference between the two phases of interest.

The purpose of this paper is to give a detailed description of the IP method and show that it is a viable way of computing the solid-liquid Gibbs free energy and construct phase diagrams. As a test case, we investigate the Lennard-Jones (LJ) model \cite{lennard-jones1924}. The remainder of the paper is organized as follows. In section II we describe the IP method in general terms. In section III we define an order-parameter that distinguishes between solid and liquid by measuring long-range order. In section IV the IP method is applied to the LJ model, and statistical and systematic errors are investigated. In section V we compare the IP method to other ways of computing Gibbs free energies and phase diagrams. The paper is completed with a summary.

\section{The interface pinning method}

\begin{figure} 
\begin{center} 
\includegraphics[width=0.7\columnwidth]{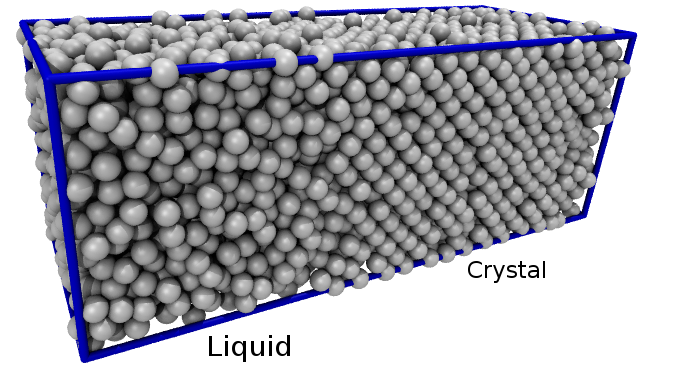}
\caption{\label{box3D} Two-phase configuration of the LJ model in a periodic orthorhombic box at a state point where the liquid is the thermodynamically stable phase while the crystal is metastable. This is an equilibrium configuration since a harmonic field biasing towards two-phase configurations have been applied. The average force exerted by the applied field on the system relates to the Gibbs free energy difference between the phases.}
\end{center} 
\end{figure}

To introduce the IP method, imagine a two-phase system \cite{woodruff1973} in a periodic orthorhombic box elongated in the $z$-direction, i.e. with box lengths $X\leq Y<Z$ (Fig. \ref{box3D}). Consider configurations of the $Np_zT$-ensemble defined as the constant temperature and pressure ensemble where the box lengths $X$ and $Y$ are fixed at values so that the crystal is unstrained, while the box length $Z$ is allowed to fluctuate in order to maintain a constant pressure. Two interfaces will form orthogonal to the long axis. This orientation will minimize the interface area and thereby the interface Gibbs free energy $G_i$. The system is only barostated in the $z$-direction since the surface tension will add an {\it a priori} unknown pressure to the $p_x$ and $p_y$ components of the pressure tensor \cite{kirkwood1949}. (It is worth noting that an orthorhombic box is not a requirement. The angle between the box vectors $\vec{X}$ and $\vec{Y}$ may differ from $90^\circ$, 
but should then be kept constant at an angle not does not strain the crystal).

\begin{figure} 
\begin{center} 
  \begin{center} 
  \includegraphics[width=0.8\columnwidth]{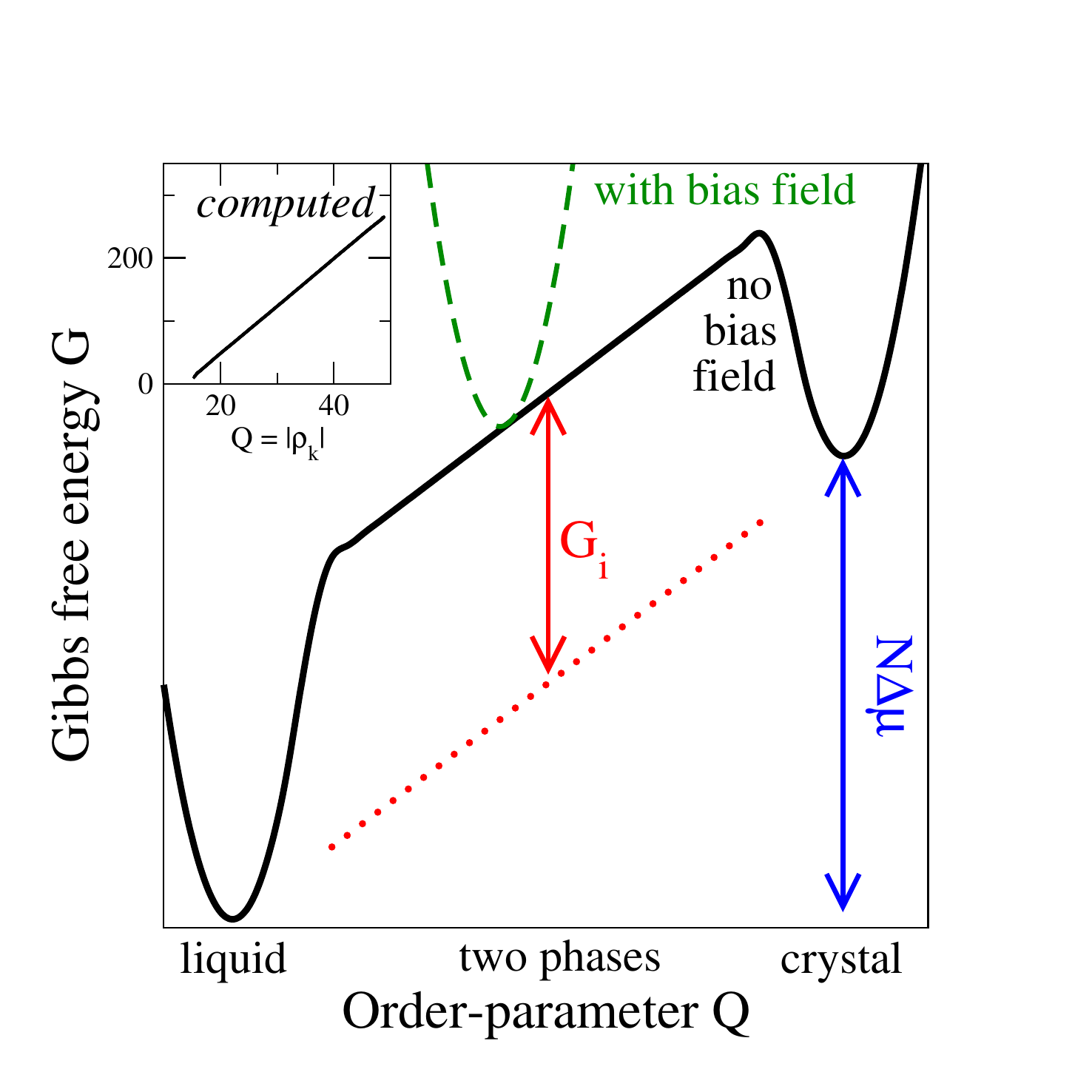} 
  \end{center} 
  \caption{\label{sketch} Sketch of Gibbs free energy $G(Q)$ (solid; black) along an order-parameter $Q$ that measure the amount of crystalline particles in a system similar to the on shown on Fig. \ref{box3D}. The liquid have the lowest $G$ and it is thus the thermodynamically stable phase while the crystal is a metastable phase. The double arrows in the center of the figure indicate the interface contribution $G_i$ to $G(Q)$. The double arrow on the right hand side of the figure indicates the total change of $G$ when moving from one phase to the other. The dashed curve indicates the Gibbs free energy $G'(Q)$ of a system where a harmonic external field has been applied to stabilize two-phase configurations. The inset shows a computed $G(Q)$ for the LJ model in the two-phase region ($N=5120$; $T=0.8$; $p=1.5$; computed using umbrella sampling \cite{frenkel2002}).} 
\end{center} 
\end{figure}

Assume that the system is sufficiently large so that the central regions of the pure phase slabs exhibit bulk properties up to some arbitrary threshold values. Particles can then be labeled either $s=\textrm{[solid]}$, $l=\textrm{[liquid]}$ or $i=\textrm{[interface]}$ and the total number of particles can be written as
\begin{equation}
   N=N_s+N_l+N_i.
\end{equation}
Let $\mu_s$ and $\mu_l$ be the chemical potential of the solid and the liquid, respectively. The total Gibbs free energy of the two-phase system is then 
\begin{equation} 
 G=N_s\mu_s + N_l\mu_l + G_i. 
\end{equation} 
as sketched on Fig. \ref{sketch}. When the relative position of the interfaces change within the two-phase regime, $G_i$ and $N_i$ can be replaced by an constant if ``wiggles'' can be neglected (wiggles \cite{troster2005,troster2005II} are discussed later in the paper). Thus combining the last two equations gives
\begin{equation}\label{GofN}
G=N_s \Delta \mu + \textrm{const.}
\end{equation} 
where $\Delta \mu \equiv \mu_s-\mu_l$. Throughout the paper we let ``$\Delta$'' denote ``$\textrm{[solid]}-\textrm{[liquid]}$'' and let ``$\textrm{const.}$'' refer to an constant.

\subsection{Harmonic field biasing towards two-phase configurations}

\begin{figure} 
\begin{center} 
  \includegraphics[width=0.8\columnwidth]{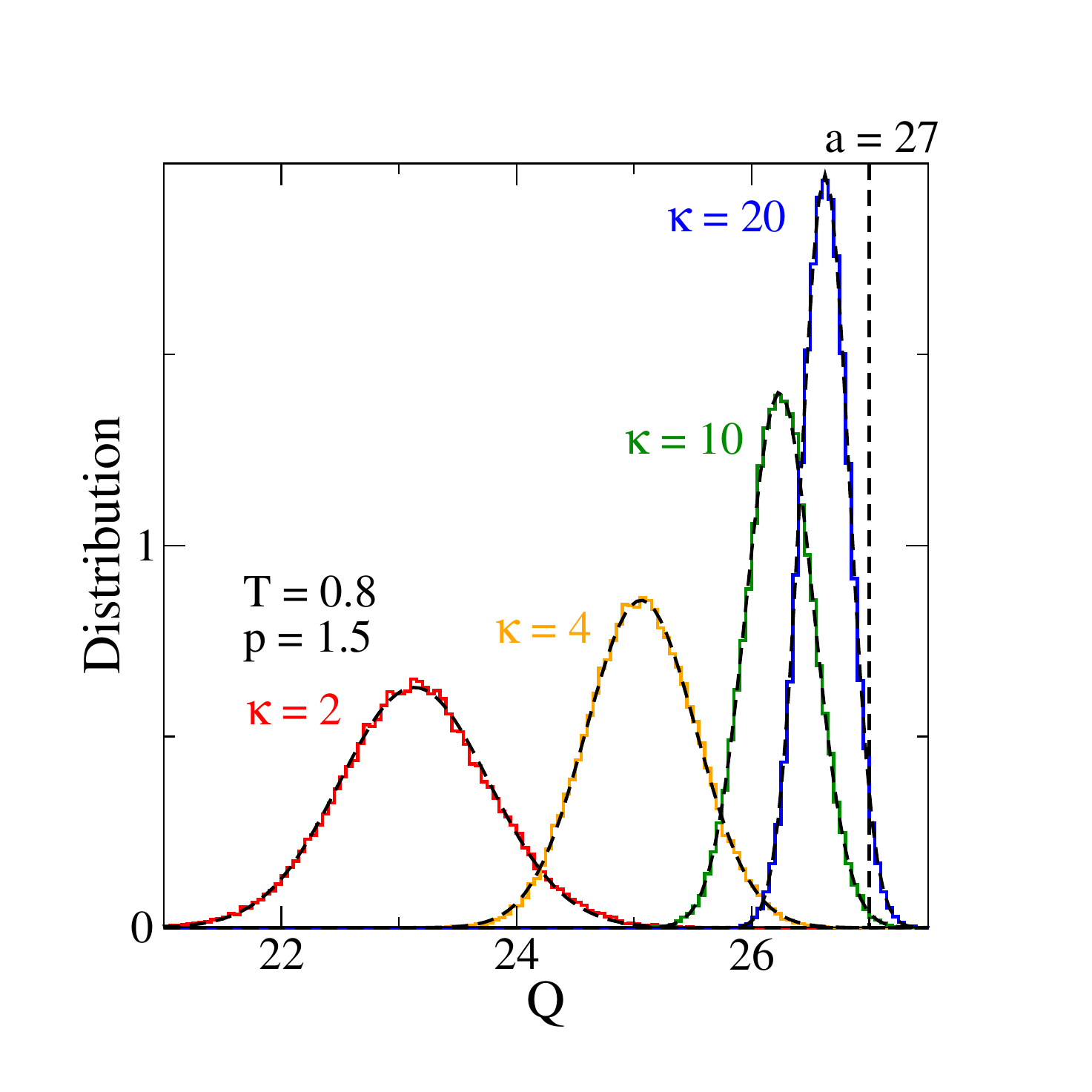}
\caption{\label{Q}  $P'(Q)$ distribution of a two-phase system where the relative interface position have been pinned with a harmonic field (LJ model; $T=0.8$; $p=1.5$; $r_c=2.5$; $t_{\rm sim}=4000$). Four values of spring constants have been used, $\kappa=\{2,4,10,20\}$ respectively. Fluctuations of the order-parameter $Q$ follow Gaussian statistics (dashed lines; Eq. (\ref{PQ})). The liquid is the thermodynamically stable phase at this state point and the average value of $Q$ is pulled by the system to values below the anchor point of $a=27$.}
\end{center} 
\end{figure}

To maintain the system in configurations having two phases, i.e. ``pinning the interfaces'', we apply a harmonic field that couples to an order-parameter which relates to the amount of crystal particles in the simulation box. To this aim we introduce a global order parameter $Q(\mathbf{R})$ where $\mathbf{R}=\{\mathbf{r}_1,\mathbf{r}_2,\ldots,\mathbf{r}_N\}$ is the configuration of particles (for simplicity we assume that particles are point-like). Let $Q_s$ and $Q_l$ be the average values of $Q(\mathbf{R})$ when the system is completely solid or liquid, respectively (at a given pressure $p$ and temperature $T$). We define $Q$ so that it has a linear dependency on the amount of solid particles in a two-phase state when additional degrees of freedom are integrated out (such as phonon vibrations in the slabs of the pure phases):
\begin{equation}\label{Qlin} 
 Q = \frac{N_s}{N}Q_s+\frac{N_l}{N}Q_l+\frac{N_i}{N}Q_i =N_s\frac{\Delta Q}{N}+\textrm{const.}
\end{equation} 
where $\frac{N_i}{N}Q_i$ is a constant contribution from interface particles.
Let $U(\mathbf{R})$ be the energy of the unperturbed system, and
\begin{equation} 
 U'(\mathbf{R})=U(\mathbf{R})+\frac{\kappa}{2}[Q(\mathbf{R})-a]^2,
\end{equation} 
be the energy when a ``spring-like'' harmonic field is applied. We refer to the field parameters $a$ and $\kappa$ as the anchor point and the spring constant, respectively. The Gibbs free energy along the $Q$ coordinate when the biasing field applied is (dashed line on Fig. \ref{sketch})
\begin{equation}
 G'(Q)=G(Q)+\frac{\kappa}{2}[Q-a]^2.
 \label{reweight}
\end{equation}
To give an expression for the probability distribution of $Q$ we use that $P'(Q)\propto \exp(-G'(Q)/k_BT)$. By insertion of Eq. (\ref{GofN}) and elimination of $N_s$ with Eq. (\ref{Qlin}) we get that,
\begin{equation} \label{PQ} 
P'(Q) = \sqrt{\frac{\kappa}{2\pi k_BT}} \exp\left(-\frac{\kappa}{2k_BT}\left[Q-a+\frac{N\Delta\mu}{\kappa\Delta Q}\right]^2\right),
\end{equation} 
in the two-phase regime. This distribution is shown for the LJ model in Fig. \ref{Q}.

\subsection{Computing $\Delta\mu$ from the average force exerted by the applied field on the system}

The chemical potential difference $\Delta\mu$ can be computed from the average force
\begin{equation}
F^{\rm field}=-\kappa[\langle Q\rangle'-a]
\end{equation}
that the field exerts on the system (along the $Q$ coordinate). When equilibrium is established the relative position does not change up to thermal fluctuations and $F^{\rm field}=-F^{\rm system}$ where $F^{\rm system}=-\frac{\partial G}{\partial Q}$. By applying the chain rule $\Delta\mu=\frac{\partial G}{\partial N_s}=\frac{\partial G}{\partial Q}\frac{\partial Q}{\partial N_s}$ then
\begin{equation}\label{dmu}
 \Delta\mu = -\frac{\kappa\Delta Q}{N}[\langle Q\rangle'-a]
\end{equation}
where $\frac{\partial Q}{\partial N_s}=\frac{\Delta Q}{N}$ is obtained from Eq. (\ref{Qlin}). Alternatively, a statistical mechanical deviation of this is possible by isolating $\Delta\mu$ from the average of the $P'(Q)$ distribution (Eq. (\ref{PQ})).

\subsection{Computing coexistence state points with the Newton-Raphson method for finding roots}

Coexistence state points are defined as $\Delta\mu(p,T)=0$ and may be computed efficiently using the Newton-Raphson algorithm for finding roots. The required derivatives of $\Delta\mu$ along isobars and isotherms
are given by the standard thermodynamic expressions
\begin{equation}
\left.\frac{\partial[\Delta\mu]}{\partial p}\right|_T=\Delta v 
\end{equation}
 and
\begin{equation}
\left.\frac{\partial[\Delta\mu]}{\partial T}\right|_p=-\Delta s
\end{equation}
where 
\begin{equation}
 \Delta s=\frac{\Delta u + p\Delta v - \Delta\mu}{T}.
\end{equation}
In these relations, $\Delta v$, $\Delta s$, and $\Delta u$ are changes in specific volume,
entropy, and energy, respectively. Thus coexistence points can be computed by iteration of
\begin{equation}\label{pip1}
p^{(i+1)} = p^{(i)} - \frac{\Delta\mu^{(i)}}{\Delta v^{(i)}}
\end{equation}
or
\begin{equation}\label{Tm_est}
T^{(i+1)} = T^{(i)}+\frac{\Delta\mu^{(i)}}{\Delta s^{(i)}}.
\end{equation} 
Iterations are continued until the computed $\Delta\mu$ is zero within the statistical error.

\subsection{Algorithm for computing coexistence state points}
To conclude this section, we give an algorithm for computing coexistence state points in the phase diagram:
First, select a pressure $p$ and temperature $T$ for the initial set of simulations. Then,
\begin{description} 
 \item[i] Construct a crystal configuration in an elongated orthorhombic box;
 \item[ii] Determine the lattice constants of the unstrained crystal by preforming $NpT$ simulation where the box lengths $X$, $Y$ and $Z$ are allowed to fluctuate independently to maintain constant pressure;
 \item[iii] Compute $Q_s$ and $v_s$ in an $Np_zT$ simulation of the unstrained crystal;
 \item[iv] Construct a liquid configuration in an elongated orthorhombic box having the same box lengths $X$ and $Y$ as the unstrained crystal;
 \item[v] Compute $Q_l$ and $v_l$ in a $Np_zT$ simulation of the liquid;
 \item[vi] Construct a two-phase configuration having the same box lengths $X$ and $Y$ as the unstrained crystal;
 \item[vii] Compute $\langle Q\rangle$ in an $Np_zT$ simulation of the two-phase system with an interface pinning $\frac{\kappa}{2}(Q(\mathbf{R})-a)^2$ field applied;
 \item[viii] Calculate $\Delta \mu$ using Eq. (\ref{dmu});
 \item[ix] If $\Delta \mu$ is non-zero within the statistical error, repeat steps i-ix at the pressure given by Eq. (\ref{pip1}) or the temperature given by Eq. (\ref{Tm_est}).
\end{description}
We note that an algorithm only involving two-phase simulations can be designed, since a two-phase simulation contains information about the bulk properties of the liquid and the crystal. In this paper we choose to use the above algorithm for practical reasons.

\section{Translational order parameter}

To utilize the method we need to define an order parameter $Q(\mathbf{R})$ that distinguishes between the phases of interest. 
Unlike liquids, crystals have long-ranged translational order and the collective density field may be used to define $Q(\mathbf{R})$:
\begin{equation}
 Q(\mathbf{R})=|\rho_{\bf k}|
\end{equation}
where 
\begin{equation}\label{eq_rho_k}
 \rho_{\bf k} = N^{-\frac{1}{2}} \sum_{j=1}^N \exp ( -i{\bf k}\cdot{\bf r}_j ),
\end{equation}
and ${\bf k}=(2\pi n_x/X,2\pi n_y/Y,0)$. The integers $n_x$ and $n_y$ are chosen so that the wave vector ${\bf k}$ corresponds to a Bragg peak. This will maximize the contrast between the liquid and the crystal. The ${\bf k}$ vector is in the $xy$-plane ($n_z=0$) since $Z$ fluctuates in the $Np_zT$-ensemble. The factor $N^{-\frac{1}{2}}$ ensures scale invariance for the average liquid value, $Q_l\propto1$, while the intensity of a (single) crystal will scale as $Q_s\propto N^{\frac{1}{2}}$. For two-phase configurations $Q$ will scale linearly with the amount of crystal particles (fulfilling Eq. (\ref{Qlin})) since the $\rho_{\bf k}$-argument of the crystal slab and liquid slabs is independent of each other.
We note that this order-parameter may be problematic in the supercooled regime, since a crystal can lower $|\rho_{\bf k}|$ by introducing a long-wave length displacement of particles. This can be avoided by using a different order-parameter, e.g. the Steinhard $Q_6$ order-parameter \cite{steinhardt1983}. This was done in Ref. \cite{pedersen_short}. We choose to use $|\rho_{\bf k}|$ as order-parameter since it is conceptually appealing, simple and general applicable.

Equilibrium trajectories can be constructed using standard Monte Carlo sampling or standard Molecular Dynamics simulations \cite{frenkel2002}. For the latter, forces exerted on particles by the external field have to be evaluated:
The force acting on particle $j$ is
\begin{equation}\label{Fj}
{\bf F}_j'={\bf F}_j-\kappa ( |\rho_{\bf k}| - a ) \nabla_j |\rho_{\bf k}|
\end{equation}
where ${\bf F}_j$ is the force without external field, and
\begin{equation}\label{Fj_nabla}
 \nabla_j |\rho_{\bf k}| = -{\bf k} \frac{\Re[\rho_{\bf k}]\sin({\bf k}\cdot{\bf r}_j) + \Im[\rho_{\bf k}] \cos( {\bf k}\cdot{\bf r}_j ) }{|\rho_{\bf k}|\sqrt{N}}
\end{equation}
where $\Re[\rho_{\bf k}]=\sum_{j=1}^N \cos({\bf k}\cdot{\bf r}_j)/\sqrt{N}$ and $\Im[\rho_{\bf k}]=-\sum_{j=1}^N\sin({\bf k}\cdot{\bf r}_j)/\sqrt{N}$ are the real and imaginary parts of $\rho_{\bf k}$ respectively.
Forces can be computed with an efficient $O(N)$ scaling algorithm although the force on particle $j$ depends on the position of all of the particles (this typically result in a $O(N^2)$ scaling algorithm). This is done in two $O(N)$ steps: i) compute $\rho_{\bf k}$ using Eq. (\ref{eq_rho_k}) and ii) compute particle forces using Eqs. (\ref{Fj}) and (\ref{Fj_nabla}).
Monte Carlo sampling involves evaluation of the energy change $\delta U'$ when a particle is moved or the box length $Z$ is changed:
\begin{equation}
 \delta U' = \delta U + \frac{\kappa}{2}\delta|\rho_{\bf k}|^2 - \kappa a\delta|\rho_{\bf k}|
\end{equation}
where $\delta|\rho_{\bf k}|^2 = |\rho_{\bf k}^{\rm try}|^2-|\rho_{\bf k}^{\rm current}|^2$ and $\delta|\rho_{\bf k}| = |\rho_{\bf k}^{\rm try}|-|\rho_{\bf k}^{\rm current}|$. These changes may be computed by evaluating $\delta\rho_{\bf k}=\rho_{\bf k}^{\rm try}-\rho_{\bf k}^{\rm current}$ if the current value of $\rho_{\bf k}=\rho_{\bf k}^{\rm current} $ is stored. Moving particle $j$ yields 
\begin{equation}
  \delta\rho_{\bf k} = [\exp ( -i{\bf k}\cdot{\bf r}_j^{\rm try} ) -  \exp ( -i{\bf k}\cdot{\bf r}_j^{\rm current} )]/\sqrt{N}.
\end{equation}
Thus computing $\delta\rho_{\bf k}$ only involves information about particle $j$ allowing for efficient computations. When the box length $Z$ is change $\delta\rho_k=0$ and $\delta U' = \delta U$ since $\mathbf{k}$ is perpendicular to the $z$-direction.

\section{Solid-liquid computations of the Lennard-Jones model}

As a test case we apply the IP method to compute solid-liquid Gibbs free energy differences of the LJ model \cite{lennard-jones1924}. LJ interactions are truncated and shifted: $U=\sum^N_{i>j}u_{ij}$ where
$
u_{ij}=4\varepsilon ([\frac{\sigma}{r_{ij}}]^{12}-[\frac{\sigma}{r_{ij}}]^{6})-4\varepsilon ([\frac{\sigma}{r_{c}}]^{12}-[\frac{\sigma}{r_{c}}]^{6})
$ 
for $r_{ij}<r_c$ and zero otherwise. LJ units are used throughout the paper: $\varepsilon=\sigma=m=k_B=1$. Two truncation distances are considered: $r_c=2.5$ and $r_c=6$. The first choice of 2.5 is the standard truncation of the LJ model. The latter choice of 6 is a better approximation of the full LJ model ($r_c\rightarrow\infty$). Molecular dynamics simulations are perform using the LAMMPS software package \cite{lammps}. The $\rho_{\bf k}$-field was implemented into the package. The Parrinello-Rahman barostat is used \cite{parrinello81} with a time constant of 8 Lennard-Jones time units together with a
Nos{\'e}-Hoover \cite{nose1984,hoover1985} thermostat with a time constant of $\tau_{\rm NH}=4$.
Trajectories are evaluated using a time step of 0.004.

\begin{figure} 
\begin{center} 
  \includegraphics[width=0.8\columnwidth]{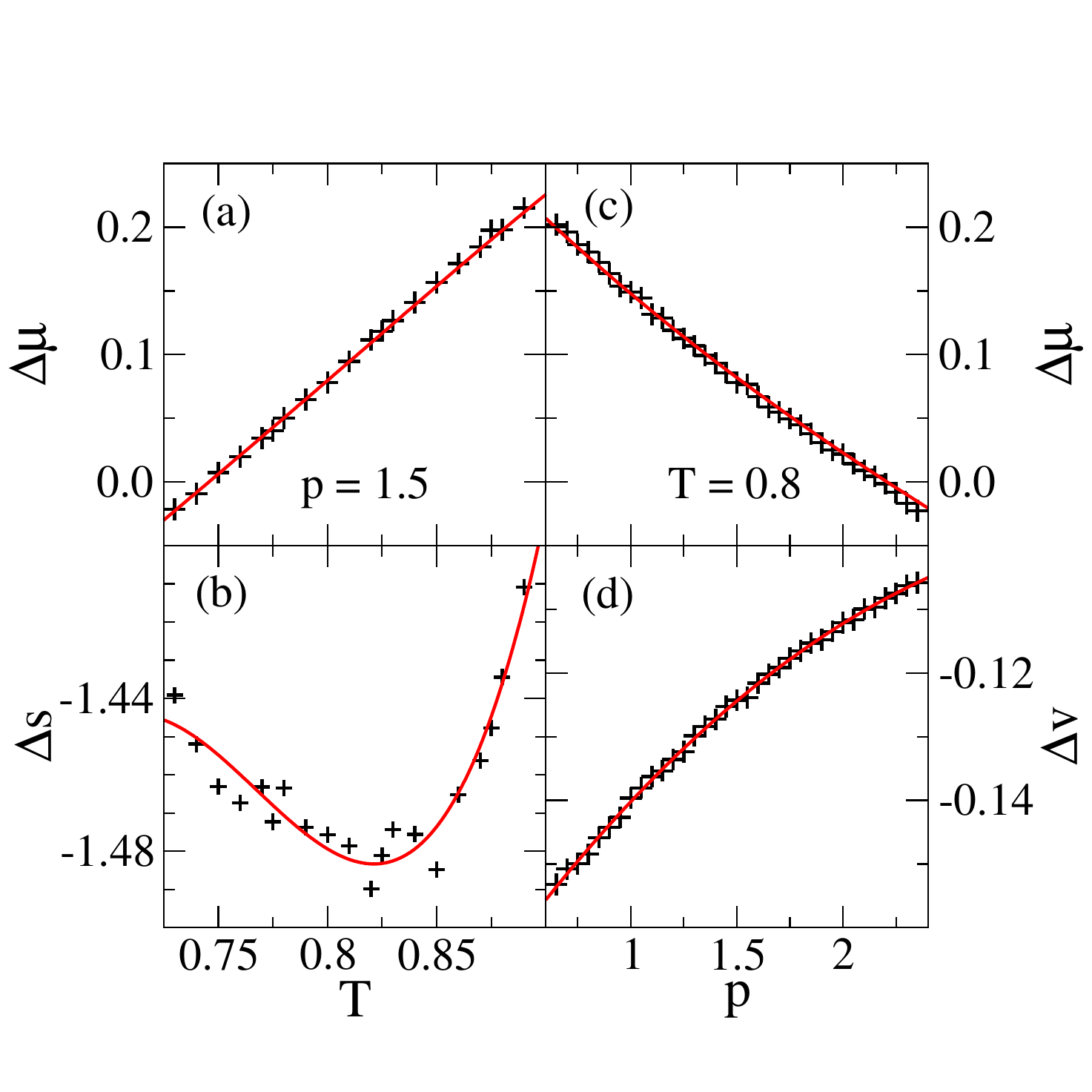}
  \caption{\label{consist} Panels (a) and (c) show the Gibbs free energy difference between solid and liquid computed with IP method ($+$) and thermodynamic integration (lines). Panels (b) and (d) show the specific entropy and the specific volume, respectively. The lines on the lower panels are cubic polynomial fits. The lines in the upper panels are computed by integration of the fits. The integration constant is chosen to give the best overall agreement ($r_c=2.5$).}
\end{center}
\end{figure}

As an example, we compute $\Delta\mu$ at $p=1.5$ and $T=0.8$ ($r_c=2.5$) as described in the following. First, a crystal structure of 8$\times$8$\times$20
face centered cubic unit cells  ($N=5120$) is constructed and
simulated for $t_{\rm sim}=4000$. All box lengths are allowed to fluctuate in order to
determine the lattice constants of the unstrained crystal giving box lengths of $X=Y=12.92$.
The unstrained crystal is then simulated for
$t_{\rm sim}=4000$ in the $Np_zT$ ensemble, and $Q_s=55.04$ $(n_x=16,n_y=0)$ and the average partial volume $v_s=1.052$ is
recorded. Next, a liquid configuration is constructed by melting the
crystal in a constant volume simulation by simulating at a high temperature ($T=5$). The
$Np_zT$-ensemble (using $X=Y=12.92$) of the liquid is then simulated for
$t_{\rm sim}=4000$. $Q_l=0.93$ and an average specific
volume of $v_l=1.177$ is recorded. Then, a two-phase
configuration is constructed by performing a high temperature
constant volume simulation where particles at $z<Z/2$ are kept at
their crystal positions using harmonic springs anchored at crystal
sites. The box volume is set in between that of the crystal and the liquid by scaling the box length $Z$. 
The $Np_zT$-ensemble with a harmonic bias-field ($a=27$;
$\kappa=10$) is then simulated for $t_{\rm sim}=40000$ to compute
$\langle Q\rangle'= 25.246$. Eq. (\ref{dmu}) yields a chemical
potential difference of $\Delta\mu = 0.080$. 
A configuration from this last simulation is shown in Fig. \ref{box3D}.
The two upper panels in Fig. \ref{consist} show $\Delta\mu$ along the $p=1.5$ isobar and the $T=0.8$ isotherm, respectively ($r_c=2.5$). The solid lines are computed with thermodynamic integration of $\Delta s$ and $\Delta v$, respectively (shown in the lower panels). The agreement is excellent.

Next, we use the Newton-Raphson method along isotherms to compute coexistence state points. As an example we computed the $T=0.8$ coexistence pressure from the state point at $p^{(1)}=1.5$. Eq. \ref{pip1} provides pressures of $p^{(i)} =\{2.141,2.189,2.185(2)\}$.
In the last iteration the estimated chemical potential difference is zero within the statistical error, $\Delta\mu=0.0001(2)$ (numbers in parentheses indicate the statistical errors on the last digit). 
Table \ref{tbl_ljcoex25} and \ref{tbl_ljcoex} list computed coexistence points using $r_c=6$ and $r_c=2.5$, respectively. As a consistency check, we note that the computed melting line obey the Clausius-Clapeyron relation, $\frac{dp_m}{dT_m}=\frac{\Delta s}{\Delta v}$ (two last rows in Tables \ref{tbl_ljcoex25} and \ref{tbl_ljcoex}). 
The left-hand side of the relation is computed by central differences of the computed melting line.

\begin{table*}
\caption{\label{tbl_ljcoex25}  Solid-liquid coexistence line of the truncated LJ model ($r_c=2.5$)}
\begin{ruledtabular} 
\begin{tabular}{ l  c c c c c c c c c c c c c }
$T_m$ & 0.6    & 0.7   & 0.8   & 0.9   & 1.0   & 1.2   & 1.4    & 1.6    & 1.8    & 2.0    & 2.2    & 2.4    & 2.6  \\
$p_m$ & -0.212 & 0.928 & 2.185 & 3.514 & 4.939 & 7.921 & 11.181 & 14.632 & 18.180 & 22.007 & 26.029 & 30.050 & 34.314 \\
\hline
$p_m+p^\textrm{tail}$& -1.046 & 0.049 & 1.264 & 2.555 & 3.943 & 6.859 & 10.056 & 13.448 & 16.943 & 20.717 & 24.688 & 28.661 & 32.878 \\
$v_s$ & 1.0614 & 1.0452 & 1.0277 & 1.0110 & 0.9951 & 0.9672 & 0.9421 & 0.9202 & 0.9014 & 0.8835 & 0.8671 & 0.8530 & 0.8394 \\
$v_l$ & 1.2194 & 1.1714 & 1.1360 & 1.1080 & 1.0830 & 1.0446 & 1.0117 & 0.9838 & 0.9612 & 0.9399 & 0.9211 & 0.9043 & 0.8888 \\
$u_s$ & -5.358 & -5.156 & -4.953 & -4.742 & -4.513 & -4.020 & -3.483 & -2.907 & -2.301 & -1.663 & -0.997 & -0.315 & 0.394  \\
$u_l$ & -4.294 & -4.218 & -4.075 & -3.888 & -3.683 & -3.183 & -2.627 & -2.041 & -1.400 & -0.727 & -0.009 & 0.696  & 1.446  \\
$\Delta s$
      & -1.718 & -1.507 & -1.392 & -1.327 & -1.263 & -1.207 & -1.168 & -1.123 & -1.107 & -1.091 & -1.087 & -1.063 & -1.055 \\
$\frac{dp_m}{dT_m}$
      & -      & 12.0$^a$   & 12.9   & 13.8   & 14.7   & 15.6   & 16.8   & 17.5   & 18.4   & 19.6   & 20.1   & 20.7   & - \\
$\frac{\Delta s}{\Delta v}$ 
      & 10.9   & 11.9   & 12.9   & 13.7   & 14.4   & 15.6   & 16.8   & 17.6   & 18.5   & 19.3   & 20.1   & 20.7   & 21.4 \\

\end{tabular} 
\end{ruledtabular} 
\\ $^{a}$: $\frac{dp_m}{dT_m}$ computed by central difference of values in the first two rows.
\end{table*}

\begin{table*}
\caption{\label{tbl_ljcoex}  Solid-liquid coexistence line of the truncated LJ model ($r_c=6$)}
\begin{ruledtabular} 
\begin{tabular}{ l  c c c c c c c c c c c c c }
 $T_m$   &  0.6   & 0.7    & 0.8    & 0.9   & 1.0    & 1.2    & 1.4    &  1.6    & 1.8    & 2.0   & 2.2     & 2.4    & 2.6             \\
 $p_m$   & -0.970 & 0.132  & 1.337  & 2.629 & 4.012  & 6.930  & 10.145 &  13.549 & 17.104 & 20.857& 24.850  & 28.916 & 33.041          \\
\hline 
 $p_m+p^\textrm{tail}$ 
         & -1.030 & 0.068  & 1.270 & 2.560  & 3.940  & 6.853  & 10.063 & 13.463  & 17.014 & 20.763 & 24.753 & 28.815 & 32.937 \\
 $v_s$   & 1.0553 & 1.0400 & 1.0242 & 1.0086 & 0.9933 & 0.9661 & 0.9412 & 0.9194 & 0.9002 & 0.8827 & 0.8663 & 0.8518 & 0.8388 \\
 $v_l$   & 1.2358 & 1.1804 & 1.1425 & 1.1120 & 1.0864 & 1.0470 & 1.0127 & 0.9852 & 0.9616 & 0.9403 & 0.9208 & 0.9038 & 0.8885 \\
 $u_s$   &-6.314  &-6.125  & -5.929 & -5.722 & -5.506 & -5.037 & -4.526 & -3.972 & -3.385 & -2.768 & -2.120 & -1.451 & -0.764  \\
 $u_l$   &-5.024  &-5.008  & -4.902 & -4.755 & -4.570 & -4.106 & -3.603 & -3.021 & -2.409 & -1.762 & -1.085 & -0.379 & 0.325   \\
 $\Delta s$ 
         &-1.858 &-1.622  & -1.480 & -1.377 & -1.311 & -1.245 & -1.177 & -1.153 & -1.125 & -1.103 & -1.085  & -1.073 & -1.050 \\
$\frac{dp_m}{dT_m}$
         &  -   & 11.5$^a$ & 12.5  & 13.4 & 14.3  & 15.3  & 16.5  & 17.4  & 18.3  & 19.4  & 20.1    & 20.5  & -     \\
$\frac{\Delta s}{\Delta v}$ 
         & 10.3 & 11.6     & 12.5  & 13.3 & 14.1  & 15.4  & 16.5  & 17.5  & 18.3  & 19.1  & 20.0    & 20.6  & 21.1 \\
\end{tabular} 
\end{ruledtabular} 
\\ $^{a}$: $\frac{dp_m}{dT_m}$ computed by central difference of values in the first two rows.
\end{table*}

\subsection{Isomorph prediction of the $\rho T$ coexistence region}

\begin{figure} 
\begin{center} 
  \includegraphics[width=0.8\columnwidth]{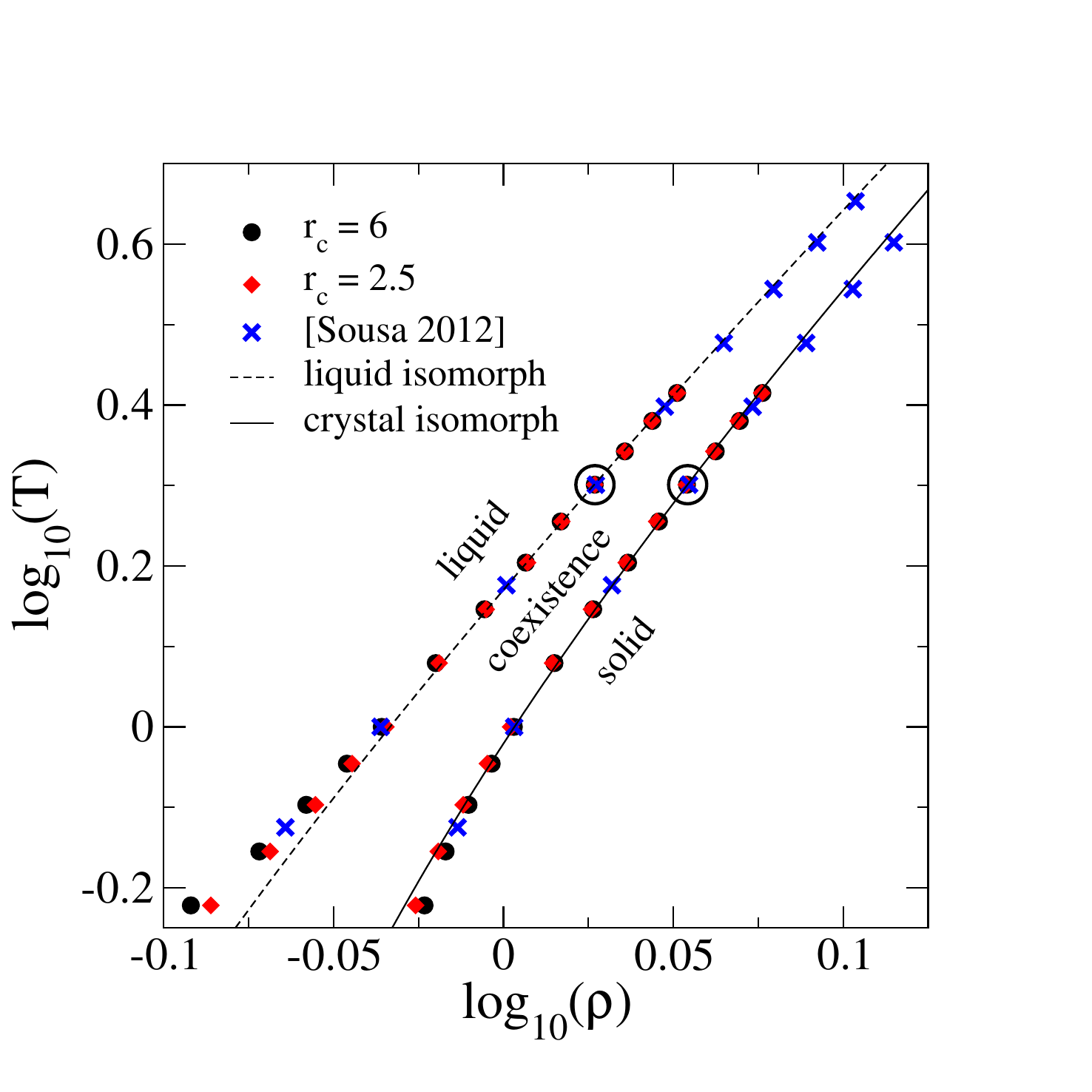}
  \caption{\label{coex_line_rhoT} Coexistence region of the Lennard-Jones model in the $\rho T$-plane on a $\log$-10 scale. Filled symbols are computed with the IP method (Tables \ref{tbl_ljcoex25} and \ref{tbl_ljcoex}). The points labeled as $\times$'s are reproduced from Reference \cite{sousa2012}. The agreement is good. At high temperatures and densities, the coexistence region is outlined by isomorphs (see text for details). The shape of the isomorphs are determined at the state points indicated by open circles (no fitting procedure was applied).} 
\end{center} 
\end{figure}

As an aside, we test a recent theoretical prediction \cite{gnan2009,schroder2011} related to the melting region in the $\rho T$-plane (Fig. \ref{coex_line_rhoT}): a large class of systems have curves in the phase diagram, referred to as ``isomorphs'' \cite{gnan2009}, where structure, dynamics and some thermodynamic properties are nearly constant. These are defined as
\begin{equation} 
  T=T_* \mathsf{h}(\rho)
\end{equation} 
where $T_*$ is the temperature at a reference state point and $\mathsf{h}(\rho)$ \cite{ingebrigtsen2012b} is a function of $\rho$ (not to be confused with the specific enthalpy $h$). 
This class of ``simple'' \cite{ingebrigtsen2012a} systems are characterized by the property that virial $W$ (the potential part of pressure) and potential energy $U$ fluctuations are strongly correlated in the NVT ensemble \cite{pedersen_prl2008,scl_I,scl_II}: if $\delta W=W-\langle W\rangle$ and $\delta U=U-\langle U\rangle$ then the correlation coefficient $R=\langle\delta W\delta U\rangle/\sqrt{\langle(\delta W)^2\rangle\langle(\delta U)^2\rangle}$ is close to unity.

For systems with inverse power-law pair interactions, $u_{ij}\propto r^{-n}$, isomorph invariance is trivial \cite{hoover1971} with $\mathsf{h}(\rho)=(\rho/\rho_*)^{n/3}$ where $\rho_*$ is the density at the reference state point. For systems with LJ pair interactions (a sum of two inverse power-laws) the scaling is approximate and reflects an effective inverse power-law of repulsive interactions \cite{pedersen_prl2010}. The apparent exponent is state point and phase dependent \cite{ingebrigtsen2012b,bohling2012}:
\begin{equation} \label{rhoh}
 \mathsf{h}(\rho) = \left[\frac{\rho}{\rho_*}\right]^4\left[\frac{\gamma_*}{2}-1\right]-\left[\frac{\rho}{\rho_*}\right]^2\left[\frac{\gamma_*}{2}-2\right]
\end{equation} 
where $\gamma_*$ is a constant that may be determined from virial-energy fluctuations in the NVT ensemble at the reference state point by using the identity $\gamma=\left.\frac{\partial \log(T)}{\partial \log(\rho)}\right|_{S_{ex}}=\frac{\langle\delta W\delta U\rangle}{\langle(\delta U)^2\rangle}$. Here it is used that excess entropy $S_{ex}$ is isomorph invariant ($\gamma_*=\gamma$ evaluated at the reference state point).

The dashed line on Fig. \ref{coex_line_rhoT} is a liquid isomorph ($\gamma_*=4.816$; $T_*=2$; $\rho_*=1/0.9403=1.064$; $R=0.991$) and the solid line is a crystal isomorph ($\gamma_*=5.517$; $T_*=2$; $\rho_*=1/0.8827=1.133$; $R=0.998$). At high temperatures and densities the coexistence region is outlined by these isomorphs. Deviations from the prediction are, however, significant at low $\rho T$. This is properly due to long-range attractive interactions not being accounted for by an effective inverse power-law. Consistent with this interpretation, the melting region of the $r_c=2.5$ truncation of pair interactions deviates from the $r_c=6$ in this part of phase space.
A scaling form of $A\rho^4+B\rho^2$ (like Eq. (\ref{rhoh})) has previously been validated by Khrapak and coworkers \cite{khrapak2011,khrapak2011b}. They used Rosenfeld's rule of additivity of melting curves \cite{rosenfeld1976} to motivate the scaling law.

\subsection{Correcting for missing long-range attractions}

To estimate the melting line of the full LJ model ($r_c\rightarrow\infty$), we applying an approximate pressure correction $p^\textrm{tail}$ that rectifies missing long-range attractions of the truncated model. To this aim we first consider the pressure correction in a simulation of solid or liquid in bulk: it is convenient to assume that the radial distribution function is constant at distance larger than the truncation. Then the correction is analytic and only depends on $\rho$ and $r_c$ \cite{frenkel2002}. Since the densities of the solid and the liquid are different, so are the corrections for the two phases. For the pressure correction of a two-phase simulation we use the average of the bulk pressure corrections:
\begin{equation}\label{ptail}
 p^\textrm{tail} = \frac{8\pi}{3}\left[v_s^{-2}+v_l^{-2}\right]\left[\frac{2}{3}r_c^{-9}-r_c^{-3}\right].
\end{equation}
Third rows in Tables \ref{tbl_ljcoex25} and \ref{tbl_ljcoex} list the corrected melting pressures ($p_m+p^\textrm{tail}$). Deviations between the corrected melting pressures when truncating at $r_c=2.5$ or $r_c=6$ are comparable to statistical error (Tables \ref{tbl_ljcoex25} and \ref{tbl_ljcoex}).

\begin{figure} 
\begin{center} 
  \includegraphics[width=0.8\columnwidth]{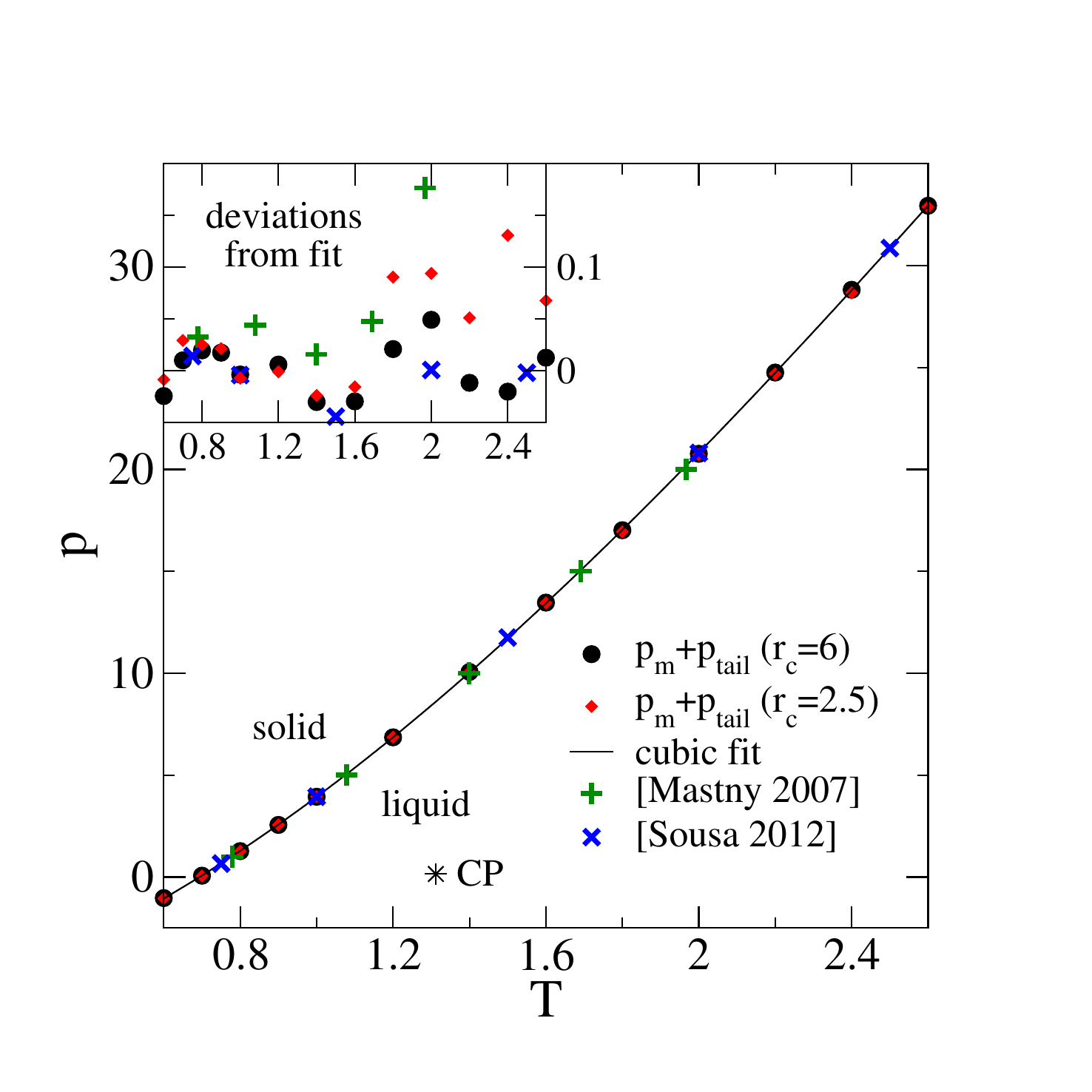}
  \caption{\label{coex_line} Coexistence line of the Lennard-Jones model in the $pT$-plane. Filled symbols are coexistence points computed with the IP method (Tables \ref{tbl_ljcoex25} and \ref{tbl_ljcoex}) corrected for truncation of long-range contributions to the pressure, Eq. (\ref{ptail}). The solid line is a cubic fit ($r_c=6$). The symbols $+$'s and $\times$'s are coexistence points reproduced from References \cite{mastny2007} and \cite{sousa2012}, respectively. The inset shows deviations from the fit. The asterisk is the gas-liquid critical point ($T_\textrm{CP}=1.31$; $p_\textrm{CP}=0.15$) \cite{potoff1998}.}
\end{center} 
\end{figure}

Computed melting points are shown on Fig. \ref{coex_line} as filled symbols. In the same figure, $+$'s and $\times$'s are coexistence points computed with other methods \cite{mastny2007,sousa2012}. The agreement is excellent. Differences are highlighted in the insert by showing deviations from a cubic fit to the computed melting points. Deviations from the results of Reference \cite{sousa2012} are within statistical error, while the melting pressure of reference \cite{mastny2007} is systematically to low by about $\Delta p\simeq0.05$ (except for one data point). Systematic errors in computed melting lines are common \cite{mastny2007} and are typically related to approximate tail corrections (like Eq. (\ref{ptail})), finite size effects or method specific systematic errors. In the following subsections we will discuss systematic and statistical errors related to the IP method.

\subsection{Statistical error}

\begin{figure} 
\begin{center} 
  \includegraphics[width=0.8\columnwidth]{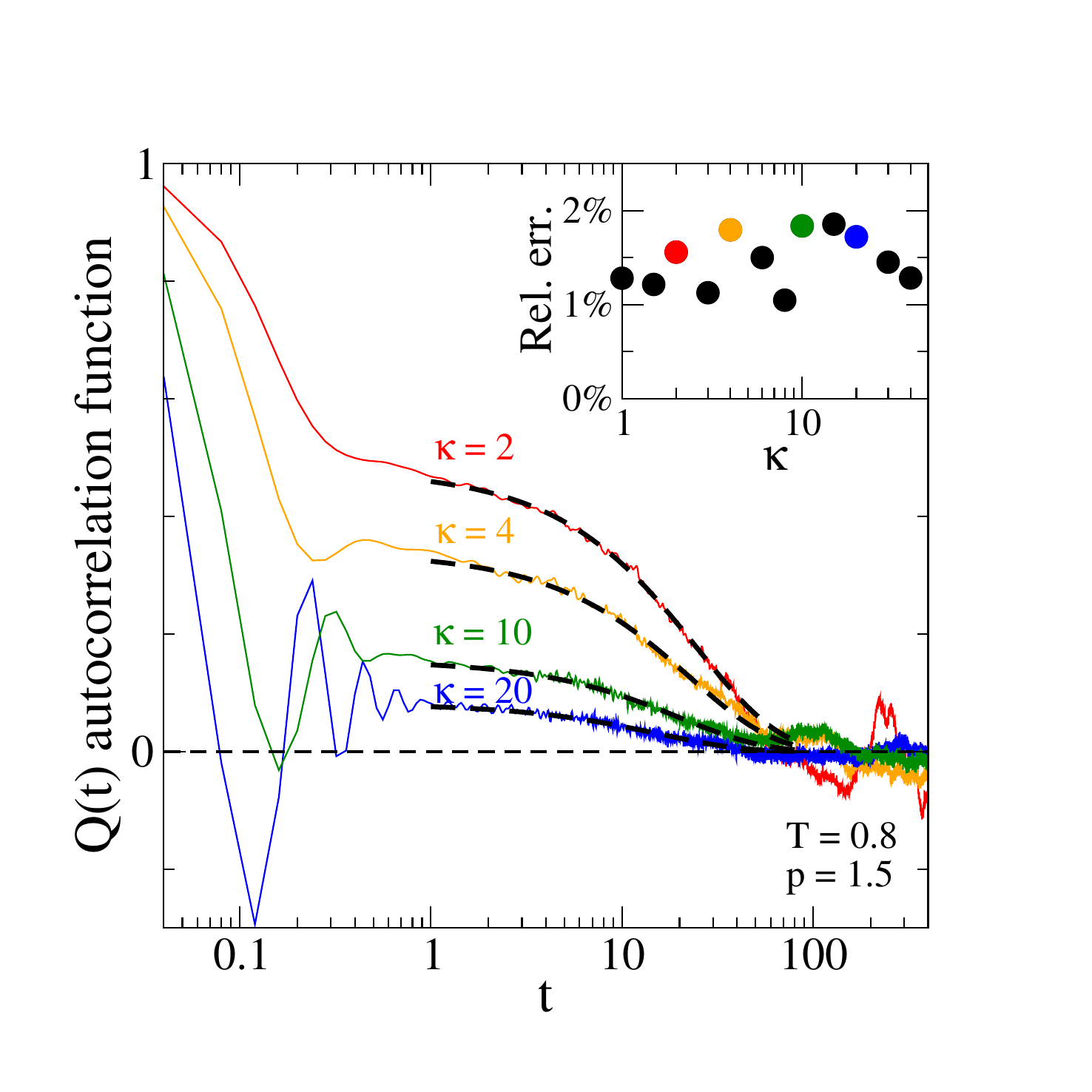}
\caption{\label{QQ} $Q$ time autocorrelation function for four spring constants $\kappa$ (same parameters as results shown on Fig. \ref{Q}). Decorrelation occurs on two timescales: i) a fast time scale related to sound waves and ii) interface movements. Dashed lines are $A\exp(-t/\tau)$ fits to the slow interface process. The inset shows the relative statistical error on the $\Delta\mu=0.080$ estimate. This error is computed by dividing runs into independent blocks \cite{flyvbjerg1989} of length $t_{\rm block}=100$ ($t_{\rm sim}=2000$).}
\end{center} 
\end{figure}

How does the statistical error of the $\Delta\mu$ estimate depend on the choice of spring constant $\kappa$?
To answer this question, we compute the $Q(t)$ autocorrelation function (using the Wiener-Khinchin theorem \cite{wiener1930} with Fast Fourier Transforms)
\begin{equation}
C(t)=\frac{\langle \delta Q(0)\delta Q(t)\rangle}{\langle (\delta Q)^2 \rangle}
\end{equation}
where $\delta Q(t) = Q(t) - \langle Q \rangle$. Fig. \ref{QQ} shows $C(t)$ for four choices of $\kappa$ ($p=1.5$; $T=0.8$; $r_c=2.5$). $C(t)$ reveals two relaxation processes that occurring on different time-scales. We assign them as follows: i) a fast process related to phonon vibrations and rearrangements of particles, and ii) a slower over-damped process related to particles moving between phases, $N_s(t)=-N_l(t)+\textrm{const}$. For the investigated $\kappa$'s the characteristic time $\tau$ for the slow process is nearly constant. $\tau$ scales as $1/\kappa$ for smaller values of $\kappa$ this timescale  (data not shown).
The relative statistical error on the computed $\Delta\mu$ is estimated by dividing runs into 20 independent blocks of length $t_{\rm block}=100>\tau$ \cite{flyvbjerg1989}. For the shown $\kappa$'s, spanning two orders of magnitude, the statistical error is independent of $\kappa$ (inset on Fig. \ref{QQ}).

\subsection{Systematic errors at small system sizes}

\begin{figure} 
\begin{center} 
  \includegraphics[width=0.8\columnwidth]{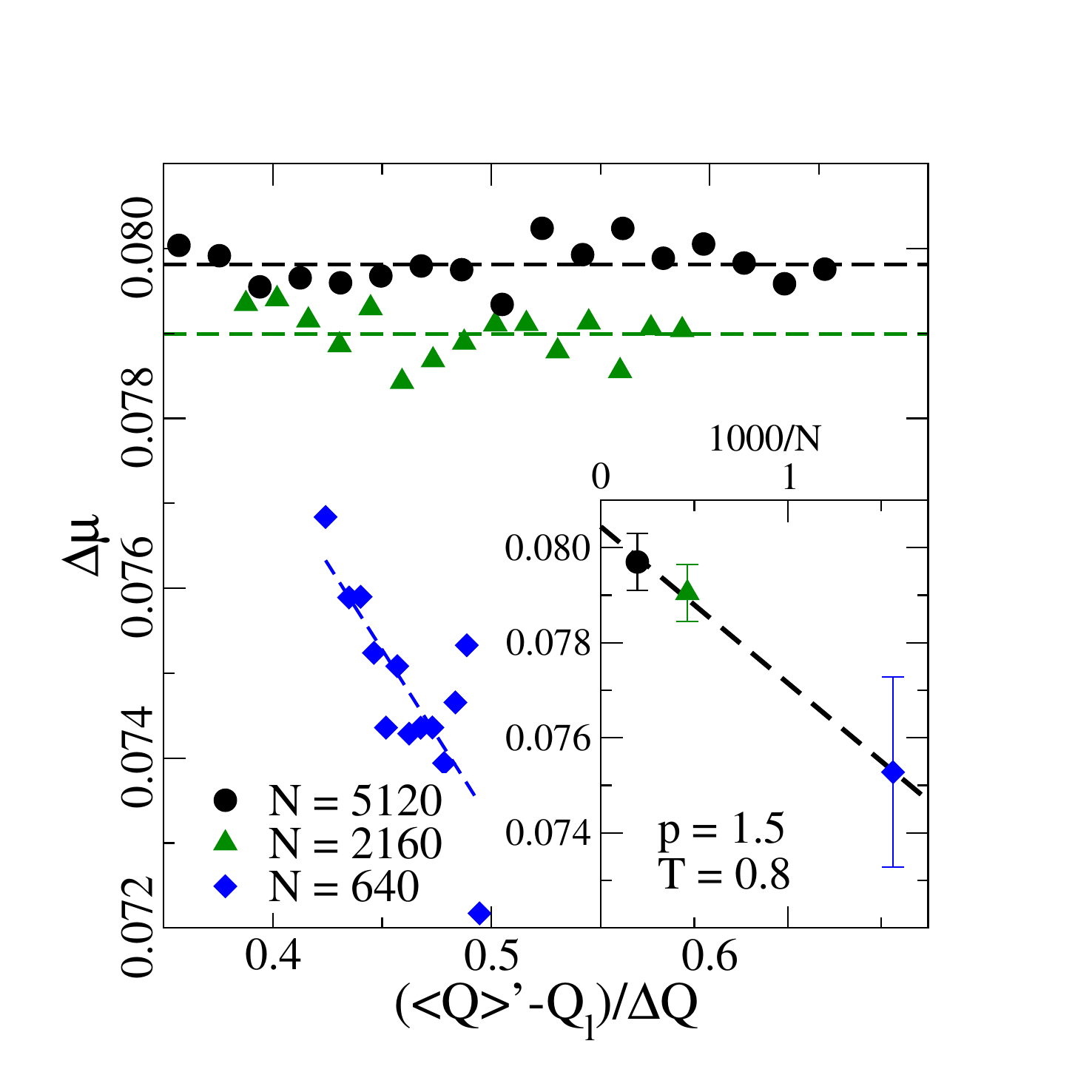}
  \caption{\label{smallsystem} Computed $\Delta\mu$ (Eq. \ref{dmu}) as a function of the ``interface distance'' defined as $(\langle Q\rangle'-Q_l)/(Q_s-Q_l)$ ($T=0.8$; $p=1.5$; $r_c=2.5$; $\kappa=10$) of three system sizes with the same geometry ($\{$4$\times$4$\times$10;6$\times$6$\times$15;8$\times$8$\times$20$\}$; $N=\{640,2160,5120\}$). The inset shows the average of the computed $\Delta\mu$'s as a function of system size. Error bars indicates the statistical error plus the variation related to interface position ($t_{\rm sim}=40000$).}
\end{center}
\end{figure}

To investigate finite size effects we computed $\Delta\mu$ at $p=1.5$ and $T=0.8$ with $r_c=2.5$ using three systems sizes with the same geometry: $N=\{640,2160,5120\}$ respectively (Fig. \ref{smallsystem}). For the smallest system sizes the computed $\Delta\mu$ depends measurably on the relative interface positions (varied by changing $a$). This effect is not seen for the two larger system sizes. The inset shows the system size dependency of the compute $\Delta\mu$. Error bars indicate statistical error plus the variation related to the positions of the interfaces relative to each other. The dashed $1/N$ line is a guide to the eye. For the largest system sizes, the error on the estimated $\Delta\mu$ is on the order of a $10^{-3}$. This correspond to an error on the computed melting temperature of about $10^{-3}$ (Eq. \ref{Tm_est}).

\subsection{Gibbs free energy dependency of the interface positions relative to each other}

\begin{figure} 
\begin{center} 
  \includegraphics[width=0.6\columnwidth]{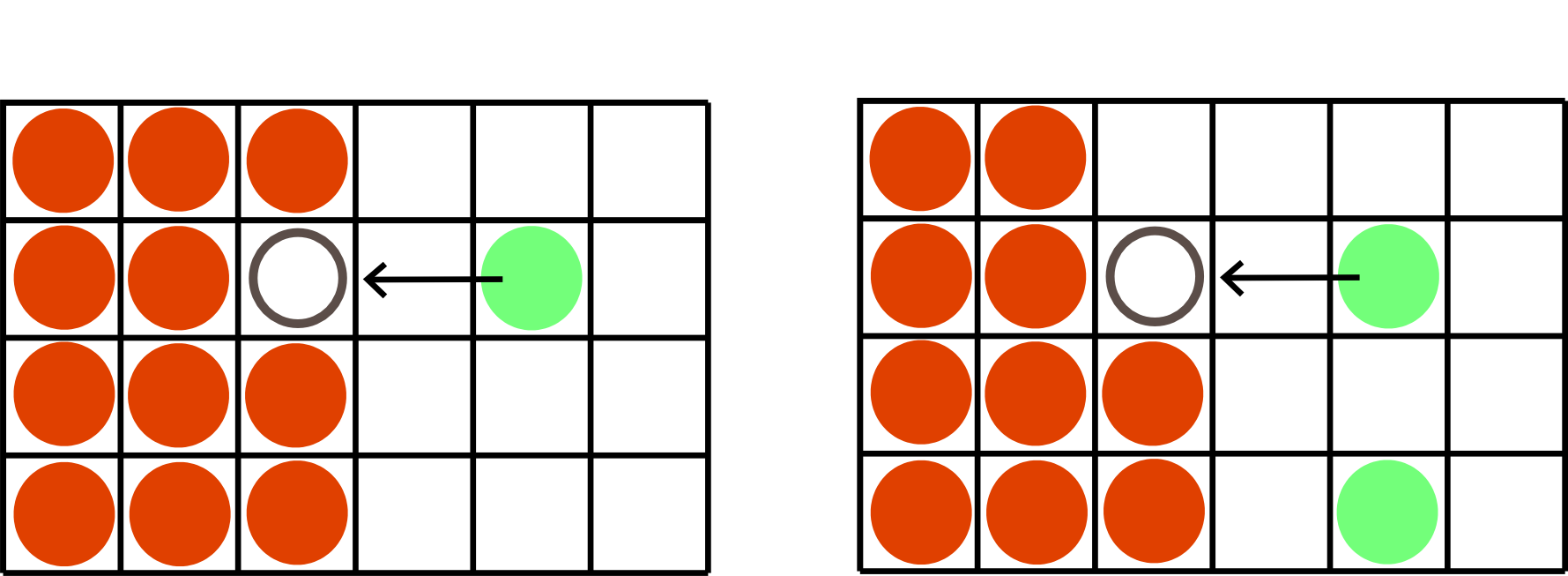}
\caption{\label{latticegas} Two two-phase configurations of a square lattice gas with attractive interactions between neighbor particles. Solid particles are colored dark gray (red) and fluid particles are colored light gray (green). The interface Gibbs free energy $G_i$ is different for the two configurations and moving a particles from one phase to the other, as indicated by the arrow, result in different changes of the energy. This result in wiggles on $G(N_s)$.}
\end{center} 
\end{figure}

\begin{figure} 
\begin{center} 
  \includegraphics[width=0.8\columnwidth]{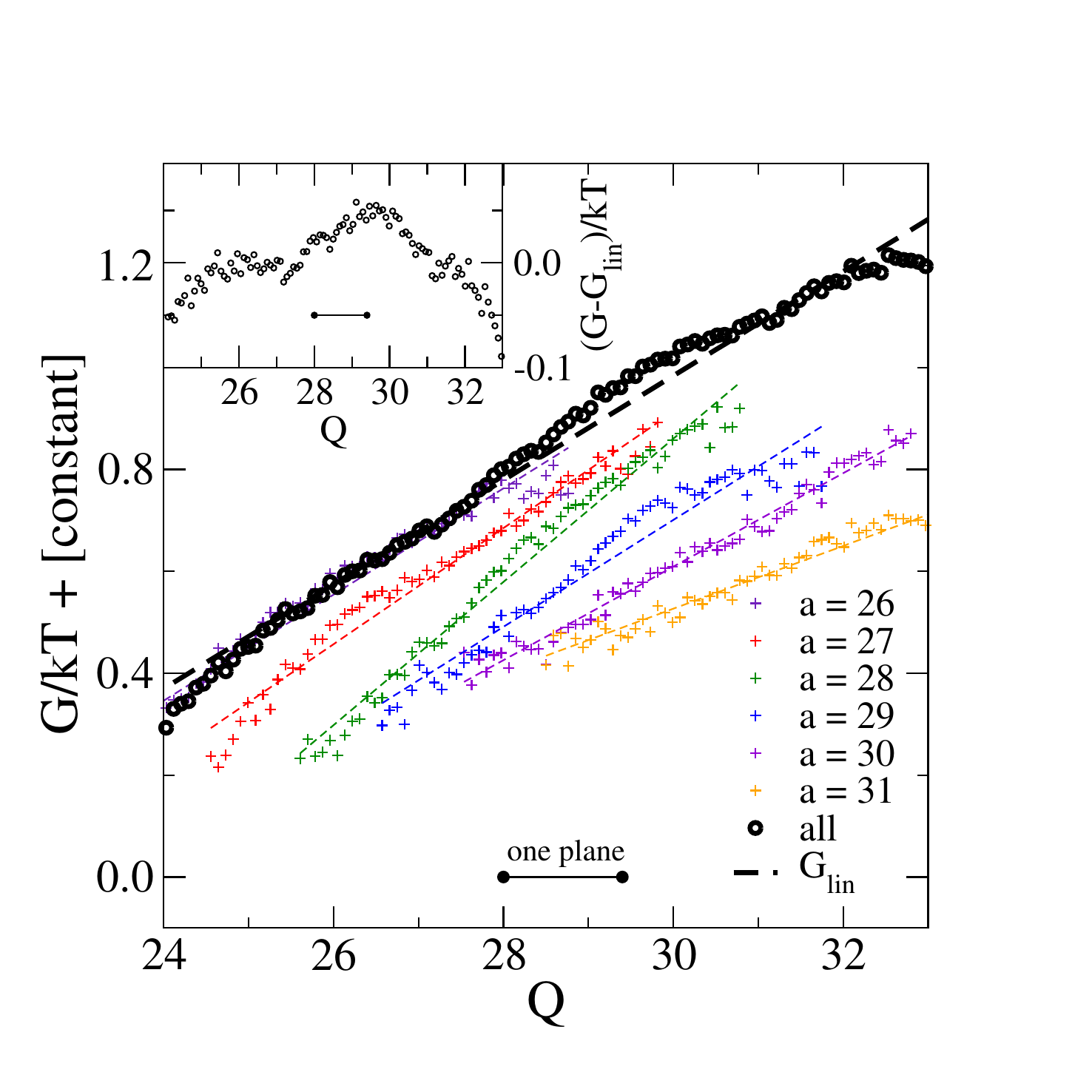}
  \caption{\label{dGdQ} Crosses show $G(Q)$ in units of $k_BT$ of six applied fields with different spring anchor points (data points are shifted horizontally for clarity). The circles show $G(Q)$ computed with umbrella sampling \cite{frenkel2002}. $G(Q)$ has a linear dependency on $Q$ (within the statistical noise).}
\end{center} 
\end{figure}

We have assumed that the Gibbs free energy in the two-phase region is independent of interface positions relative to each other.
There are, however, two effects that may spoil this assumption: i) if the distance between the interfaces is sufficiently small, particles in one (or both) phases will not have bulk properties, and ii) ``wiggles'' on $G(N_s)$ \cite{troster2005,troster2005II}. To exemplify the latter effect think of a square lattice gas with attractions between neighboring particles. Fig. \ref{latticegas} shows two two-phase configurations of this model. The interface Gibbs free energy $G_i$ is different for the two configurations resulting in wiggles of $G(N_s)$. The wriggle period on $G(Q)$ is $\Delta Q/N_z$ where $N_z$ is the number of crystal planes in the $z$-direction.
To investigate the Gibbs free energy dependency of interface positions of the LJ model, we perform simulations over a range of $a$'s with overlapping $P'(Q)$ distributions. From this $G(Q)$ is constructed with the umbrella method \cite{frenkel2002} (histogram reweighing is done with the MBAR algorithm \cite{shirt2008}). We find no wiggles (Fig. \ref{dGdQ}), but conjecture that they are hidden in the statistical noise. We emphasize that wiggles may be accounted for and do not constitute a fundamental limitation of the IP method.

\subsection{Guidelines for choosing $a$ and $\kappa$}

How should the anchor point $a$ and the spring constant $\kappa$ of the harmonic field be chosen to yield the optimal computation?
To answer this question, we note that the average distance between the two interfaces should be large as possible to ensure that slabs of the pure phases have bulk properties. This distance can be controlled by the anchor point $a$. To give a guideline for an optimal $a$, we consider the limit where $\kappa\rightarrow\infty$ or $\Delta\mu=0$. From Eq. \ref{dmu} we find that $a=\langle Q\rangle'$. The optimal value of $a$ is
\begin{equation} 
 a\simeq Q_l+\frac{\Delta Q}{2}.
\end{equation} 
The ``$\simeq$'' indicates that the interface contribution $Q_i$ has been ignored. Next, we consider the spring constant $\kappa$.
There are two arguments for choosing a stiff spring, i.e. large value of $\kappa$: i) a small $\kappa$ would not keep the system in two-phase configurations; ii) the relaxation time of interface dynamics $\tau$ scales as $1/\kappa$ for small $\kappa$'s, thus giving bad statistics. There are, however, also an argument for choosing a small $\kappa$: Interface fluctuations should span at least one crystal plane to account for wiggles on $G(Q)$. Thus, $\kappa$ should optimally be chosen to that interface fluctuation span one crystal plane. Setting the standard deviation of the $P'(Q)$ distribution (Eq. \ref{PQ}) equal to $\Delta Q/N_z$ we get:
\begin{equation} 
 \kappa \simeq k_BT \frac{N_z^2}{\Delta Q^2}
\end{equation} 
where $N_z$ is the number of crystal planes in the $z$-direction.
We emphasize that one of the conclusions of this paper is that the IP method is forgiving towards the choice of field parameters.

\section{Advantages and drawbacks of the interface pinning method}

Let us briefly review other methods for computing Gibbs free energies and phase diagrams \cite{frenkel2002,vega2008} before discussing the advantages and drawbacks of the IP method. In the moving interface approaches, discussed in the introduction of the paper, a simulation is performed of a two-phase system \cite{ladd1977,landman1986,mori1995,kyrlidis1995,agrawal2003,morris2002,hoyt2002,sibugaga2002,fernandez2006,vega2008,
weingarten2009,timan2010,pedersen2011_lwotp}. 
The thermodynamical favored phase will grow in a constant $NpT$ or $\mu VT$ simulation, allowing to locate coexistent points by changing intensive variables such as $p$, $T$ or $\mu$. An alternative is to use an indirect method where the Gibbs free energy of the pure phases is computed in separate simulations. The Gibbs free energy can be computed by Widom insertion \cite{widom1963,norman1996} or by thermodynamic integration to a state of know Gibbs free energy, e.g. an ideal gas \cite{hoover1968}, a harmonic solid \cite{hoover1971} or an Einstein solid \cite{broughton1983}. Umbrella sampling or metadynamics along a good reaction coordinate can be used to compute the Gibbs free-energy of transforming the system from one phase to the other \cite{broughton1986,angioletti-uberti2010,limmer2011,fernandez2012}. The reader is encouraged to explore Refs. \cite{frenkel2002,vega2008} and references within for more about methods for computing phase diagrams.

What are the advances and drawbacks of the IP method? First, the IP method inherits the conceptual simplicity, general applicability and ease to implement of the moving interface method. Since the solid-liquid interface is represented explicitly, simulations give information about interface properties. The surface tension may be computed by integration of pressure tensor elements \cite{kirkwood1949} or from capillary fluctuation \cite{hoyt2001}. Crystal growth rates may be computed from $Q(t)$ fluctuations (Fig. \ref{QQ}) similar to the method suggested by Briels and Tepper \cite{briels1997,tepper2002} (in the BT method, two-phase configurations are stabilized by keeping the volume constant). We leave such investigations to future publications. A disadvantage of having an explicit interface is that it is in direct contact with the solid and the liquid phases \cite{frenkel2012}. In effect, properties of the liquid and the solid slabs may not have bulk values. This is evident in the computed $\Delta\mu$'s 
of the system size with 640 particles (Fig. \ref{smallsystem}). Here $\Delta\mu$ depend systematically on $a$. This may lead to larger finite size effects compared to methods where free energies are computed in separate simulations having periodic boundaries. Another disadvantage is that a low interface mobility can result in slow dynamics (Fig. \ref{QQ}). This will result in large statistical errors or, in the worst case, it may even be difficult to reach equilibrium.

The IP method constitutes an alternative when other methods are difficult or impossible to use. 
Widom insertion \cite{widom1963,norman1996} works best for dilute systems whereas it is not a viable option for dense liquids or solids.
Thermodynamic integration to a state of known Gibbs free energy \cite{frenkel2002} is problematic when the path include additional first order transitions. This happens when a phase is surrounded by other phases in phase diagram. A reference state point can also be nontrivial to identify. Examples are quasi-crystals, liquid-crystals, plastic-crystals or other phases having a mixture of order and disorder. The IP method is versatile, and may be used to study these phases. Moreover, it can be generalized to be used with two-phase simulations of only fluid phases (gas-liquid or liquid-liquid) or only solid phases. For the latter, one of the crystals will be strained when simulating the $Np_zT$ ensemble of a two-phase configuration (if the lattice constants of the crystals are different). This can be rectified by adding the free energy of straining to the computed $\Delta\mu$. Low mobility of solid-solid interfaces could make it unfeasible to use the IP method.
Integration along a reaction path of transformation is not trivial, since it often is difficult to identify a suitable coordinate capturing the entire phase transition. As an example, the number of crystalline particles $N_s$ is not a good reaction coordinate, since the cluster of crystalline particles undergoes a geometric transition from spherical when $N_s$ is small to a slab when $N_s\simeq N/2$ (Fig. \ref{box3D}). Thus, there is at least one geometrical free energy barrier orthogonal to the $N_s$ coordinate (with a barrier height that scales with the area of the cluster: $N^\frac{2}{3}$). With the IP method, the selection of the order parameter $Q$ only has to distinguish between the two phases of interest.

\section{Summary}

In summary, we have given a detailed description of the IP method and shown that it can be used for efficient calculations of the Gibbs free energy difference between a solid and a liquid. The melting line can be computed efficiently to a high precision when the method is combined with the Newton-Raphson algorithm for finding roots. As an example, the solid-liquid coexistence line of the truncated LJ model line was computed. As an aside, it was shown that the high pressure part of the temperature-density coexistence region is outlined by isomorphs. An approximate pressure correction for rectifying truncation of pair interactions was given. Statistical errors and systematic variations were investigated. 

An important advantage of the IP method is that the solid-liquid Gibbs free energy difference is computed directly in a {\it ad infinitum} simulation at a single state point. This makes it versatile and a viable alternative when it is difficult or impossible to preform Widom insertion, thermodynamic integration or integration along a reaction path of transformation.

\section{Acknowledgments}
This work was financially supported by the Austrian
Science Fund FWF within the SFB ViCoM (F41). The author is grateful for valuable comments and suggestions from Christoph Dellago, Gerhard Kahl, Georg Kresse, Felix Hummel, Andreas Tr{\"o}ster, Emanuela Bianchi, Michael Gr{\"u}nwald, David Chandler, Patrick Varilly, David Limmer, Thomas B. Schr{\o}der, Jeppe C. Dyre and Peter Harrowell.

\bibliography{references}

\end{document}